\documentclass{llncs}
\usepackage{authblk}

\usepackage{listings}
\usepackage{color}
\usepackage{amsmath}  
\usepackage{graphicx} 
\usepackage{amssymb} 
\usepackage{algpseudocode}
\usepackage{algorithm}
\usepackage{comment}
\usepackage{hyperref}
\usepackage{orcidlink}
\usepackage{amsfonts}

\usepackage{float}
\floatstyle{plaintop}
\restylefloat{table}

\title{Post Quantum Cryptography (PQC) Signatures Without Trapdoors}
\titlerunning{Post Quantum Cryptography (PQC) Signatures Without Trapdoors}

\author{
  William J. Buchanan\orcidlink{0000-0003-0809-3523}\inst{1} 
}

\institute{
Blockpass ID Lab, Edinburgh Napier University, Edinburgh.
} 

\begin{document}
\maketitle
\begin{abstract} 
Some of our current public key methods use a trap door to implement digital signature methods. This includes the RSA method, which uses Fermat's little theorem to support the creation and verification of a digital signature. The problem with a back-door is that the actual trap-door method could, in the end, be discovered. With the rise of PQC (Post Quantum Cryptography), we will see a range of methods that will not use trap doors and provide stronger proof of security. In this case, we use hash-based signatures (as used with SPHINCS+) and Fiat Shamir signatures using Zero Knowledge Proofs (as used with Dilithium). 
\end{abstract}


\section{Introduction}
Some of our existing public key methods, such as RSA, use a trapdoor method to produce a digital signature.  The problem with a trap door is that eventually the target will find out where the catch is and not fall down the trap door. For many of our existing public key methods, we have used a trap door — a magical way to reverse an operation. The strength of the method is then the strength of the trap door method. This is the case for the RSA method, and which uses Fermat’s Little Theorem to perform the trap door. Our new post-quantum cryptography methods avoid the trapdoor technique and focus on more difficult problems. For quantum robust signatures which do not use trap doors, there are two main methods: lattice approaches with the Fiat-Shamir method \cite{fiat1986prove} for a non-interactive approach. For PQC (Post Quantum Cryptography, NIST has standardised the hash-based methods as SLH-DSA (FIPS 205, aka SPHINCS+ \cite{bernstein2019sphincs}), and the Fiat-Shamir approach with ML-DSA (FIPS 204, aka CRYSTALS Dilithium \cite{ducas2018crystals}.

\section{A trap door method}
First, let's look at a trap door method. We generate the RSA key pair with two prime numbers ($p$ and $q$), and compute the modulus ($N$) \cite{asecuritysite_58011}:

\begin{align}
N=pq
\end{align}

We then compute:
\begin{align}
\phi=(p-1).(q-1)
\end{align}

We then pick an encryption key value ($e=0x010001$) and compute:

\begin{align}
d=e^{-1} \pmod \phi
\end{align}

The public key is then $(e,N)$ and the private key is $(d,N)$. To sign the hash of a message ($H$), we create a signature with:

\begin{align}
S=H^d \pmod N
\end{align}

and then verify with:

\begin{align}
H=S^d \pmod N
\end{align}

This all works because of Little Fermat's Theorem, and which is defined as:

\begin{align}
a^{p-1} \equiv 1 \pmod {p}
\end{align}

\section{Hash-based signatures}
\begin{figure}
  \includegraphics[width=\linewidth]{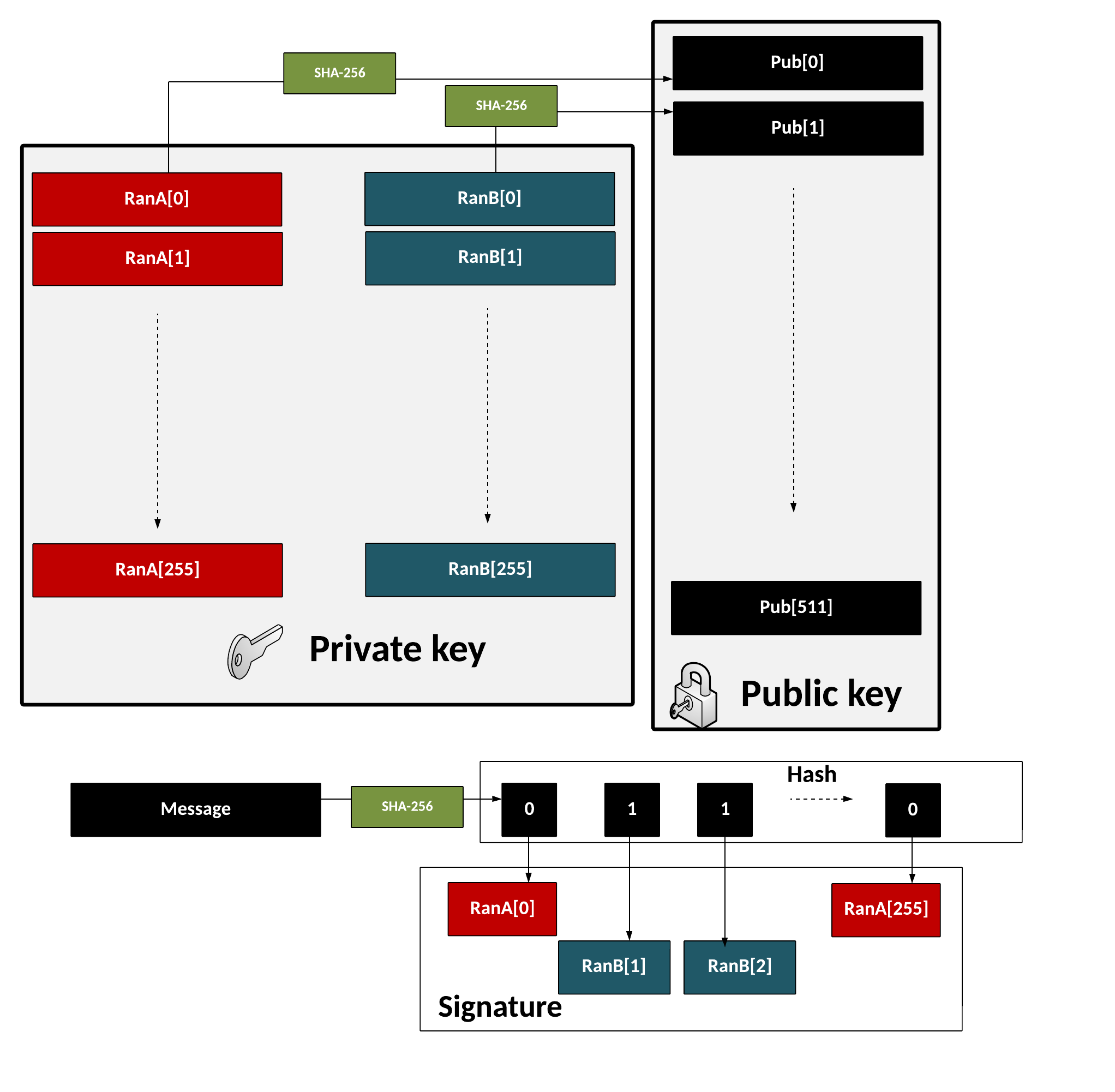}
  \caption{Lamport one-time signatures}
  \label{fig:asym}
\end{figure}

We can create a one-time signature using the Lamport approach \cite{lamport1979constructing}. With this, we generate 256 private keys for A and 256 private keys for B \cite{asecuritysite_10795}. The public key is then the hash of all of these keys (512 in total). When we sign, we take a hash of the message (giving us 256 bits of a hash). We then go through each bit, and if we have a 0, we select the associated private key from A. Otherwise, we select it from B. We then go through the bits and end up with 256 keys, either from A or B. This gives us a 256x256 signature size (8~kB). When verifying, we check that each of the hashed values is contained in the public key. This is, of course, a one-time signature, and we would have to create new keys for the next signature and republish our public key (Figure \ref{fig:asym}).

\begin{figure}
  \includegraphics[width=\linewidth]{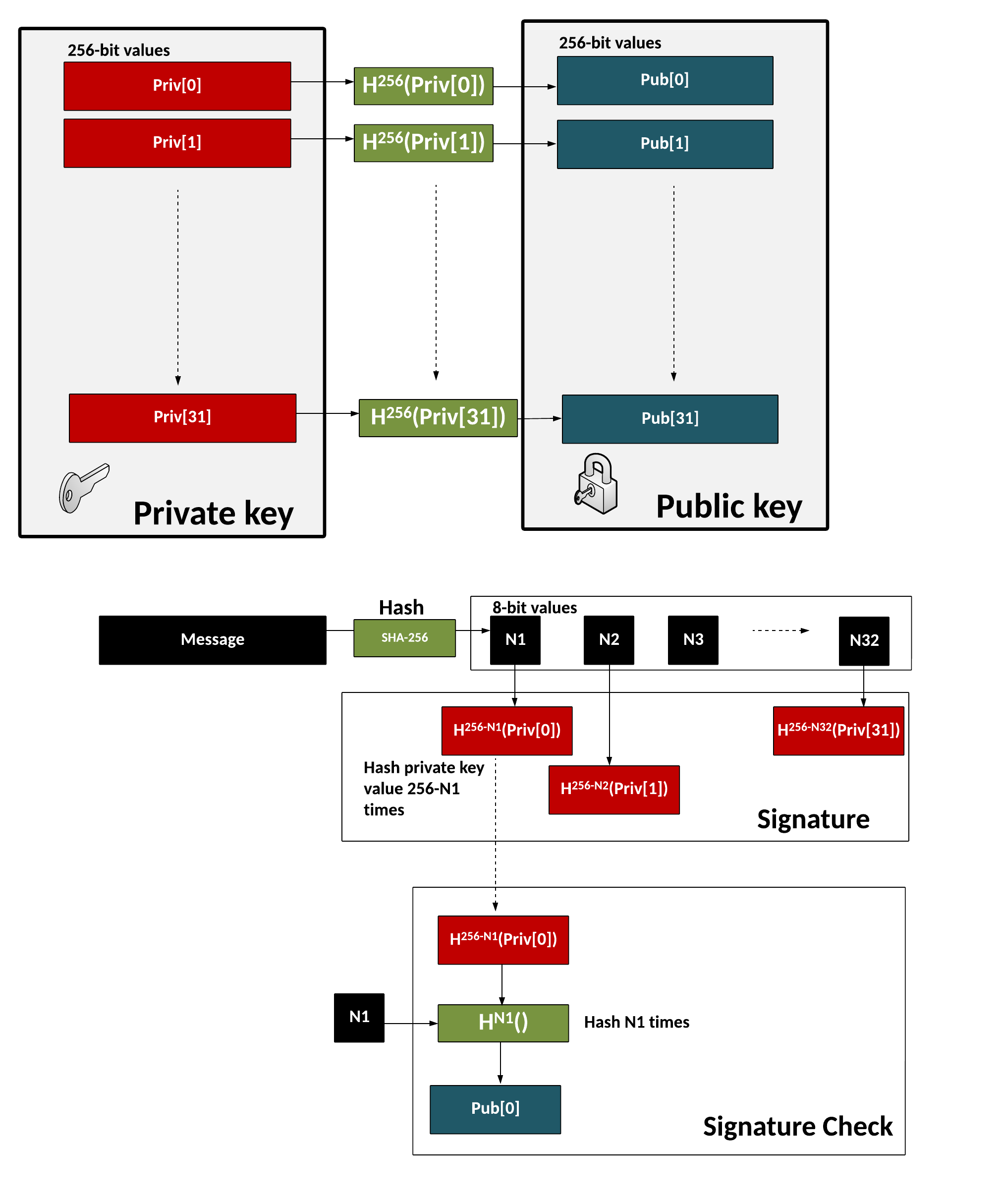}
  \caption{WOTS+ one-time signatures}
  \label{fig:wots}
\end{figure}

To improve this, we can define the WOTS+ (Winternitz one-time signature scheme) approach \cite{asecuritysite_19431,hulsing2013w}. The method is (Figure \ref{fig:wots}):

\begin{itemize}
\item Initially, we create 32 256-bit random numbers. These 32 values will be our private key.
\item We then hash each of these values 256 times. These 32 values will be our public key.
\item We now take the message and hash it using SHA-256. This produces 32 8-bit values ($N_1$, $N_2$ ... $N_{32}$).
\item For the signature, we take each 8-bit value in the hash of the message, and then hash 256-$N$ times (where $N$ is the value of the 8-bit value).
\item To prove the signature, we take the message and hash it with SHA-256, and then take each 8-bit value. We then hash the 8-bit signature value by the number of times defined by the message hash value ($N_1$, $N_2$..). The result for each operation should equal the public key value.
\item For the signature, we take each 8-bit value in the hash of the message, and then hash 256-$N$ times (where $N$ is the value of the 8-bit value).
\end{itemize}

In this way, we cut down the size of the signature and of the public key. We can then produce a Merkle Root of our public key and private key, and reduce them down to a single 256-bit value. We can then create a tree of these keys, so that we can sign more than once and have different private key, but still a single 256-bit private key and a 256-bit public key. This is the method that SPHINCS+ uses, and which has small public and private keys, but has a larger signature (around 17KB for 128-bit security):

\begin{table*}
\centering
\caption{SPHINCS+ key and cipher sizes}
\label{tab:cipher}
\begin{tabular}{p{3cm} l l l l}
\hline
Method    &   Public key size  &   Private key size  &  Signature size  \\ Security level \\
\hline
SPHINCS+ SHA-256 128-bit    &  32    &             64     &        17,088    &     1 (128-bit)\\
SPHINCS+ SHA-256 192-bit   &   48  &               96        &     35,664  &       3 (192-bit)\\
SPHINCS+ SHA-256 256-bit &     64   &             128       &      49,856   &      5 (256-bit)\\
\hline
\end{tabular}
\end{table*}

\section{Fiat Shamir signatures using Zero Knowledge Proofs}
The method for creating the lattice-based signatures is actually based on the Schnorr signature method \cite{schnorr1990efficient} for proof of identity (a Zero Knowledge Proof) and then applies the Fiat-Shamir method \cite{fiat1986prove} to make it non-interactive. Basically, it is a NI-ZKP (Non-interactive Zero Knowledge Proof) of a secret (a person’s private key).

If Bob wants to prove that he knows a secret, he creates a short random secret vector ($x$). He then creates a random matrix array of $\textbf{A}$, and an error matrix of small errors ($e_1$). His public key ($textbf{u}$) is then:

\begin{align}
u = \textbf{A}.x + e_1
\end{align}

Let’s ignore the error, as it will eventually be removed when we perform the calculations. So let’s just use:

\begin{align}
u = \textbf{A}.x
\end{align}

This value will be sent to Alice. She then can send a challenge ($c$) to Bob, and which is a random integer value. Bob then creates a random short vector of $y$, and computes:

\begin{align}
v = \textbf{A}.y + e_2
\end{align}

Let’s ignore $e_2$ again, as it will eventually be removed. Bob then computes using the Schnorr identity proof:

\begin{align}
z = c.x+y
\end{align}

Bob then sends $x$ and $v$ to Alice, and she checks if these are equal:

\begin{align}
\textbf{A}.z = c.\textbf{u} + \textbf{v}
\end{align}

If they are, then Bob has proven that he knows the value of $x$ (his private key). This works because:

\begin{align}
\textbf{A}.z = \textbf{A} (c.x+y) = \textbf{A}.c.x + \textbf{A}.y\\
c.\textbf{u}+\textbf{v} = c.\textbf{A}.x + \textbf{A}.y
\end{align}

Now, we can turn a zero-knowledge proof into a signature by using the Fiat-Shamir heuristic. With this, there is no need for Alice to send a challenge, as Bob can compute:

\begin{align}
c = \textrm{H}(v || M)
\end{align}

and where $M$ is the message, “||” is a concatenation of the values, and $H()$ is a hash function (such as with SHA-256). We can then just perform the same operations, and Alice will be able to regenerate $c$ from the values of the message and $v$, and check that:

\begin{align}
\textbf{A}.x = c.\textbf{u}+\textbf{v}
\end{align}

\section{Conclusions}
Many of our existing public key methods are based on trapdoor methods. Improved proofs of security can often be created using methods which do not have a trap door. In this case, we have outlined two core methods for creating non-trapdoor-based signatures: hash-based signatures and converting the Schnorr identity method \cite{schnorr1990efficient} into a signature using the Fiat-Shamir heuristic \cite{fiat1986prove} that converts a Zero Knowledge Proof into a digital signature method.

\bibliographystyle{IEEEtran}
\bibliography{main}

\end{document}